\documentclass[english,prd,onecolumn,superscriptaddress,nofootinbib,preprintnumbers]{revtex4}
\usepackage[latin1]{inputenc}
\usepackage{graphicx}
\usepackage{amsmath,amsthm,amssymb}

\makeatletter

\usepackage{amsfonts}
\usepackage{dcolumn}
\usepackage{hyperref}

\def\be{\begin{equation}}
\def\ee{\end{equation}}
\def\ba{\begin{eqnarray}}
\def\ea{\end{eqnarray}}
\def\bs{\begin{subequations}}
\def\es{\end{subequations}}

\newcommand{\rd}{{\rm d}}

\usepackage{color}

\makeatother

\usepackage{babel}
\makeatother
\begin{document}

\title{Revisiting chameleon gravity--thin-shells and no-shells \\
with appropriate boundary conditions}

\author{Takashi Tamaki}
\address{Department of Physics, Waseda University, Okubo 3-4-1, Tokyo 
169-8555, Japan}
\email{tamaki@gravity.phys.waseda.ac.jp}
\affiliation{Department of Physics, Rikkyo University, Toshima, Tokyo 171-8501, Japan}

\author{Shinji Tsujikawa}
\affiliation{Department of Physics, Faculty of Science, Tokyo University of Science, 
1-3, Kagurazaka, Shinjuku-ku, Tokyo 162-8601, Japan}
\email{shinji@rs.kagu.tus.ac.jp}

\begin{abstract}

We derive analytic solutions of a chameleon scalar field $\phi$ 
that couples to a non-relativistic matter in the weak gravitational 
background of a spherically symmetric body, paying particular attention 
to a field mass $m_A$ inside of the body.
The standard thin-shell field profile is recovered 
by taking the limit $m_A r_c \to \infty$, where
$r_c$ is a radius of the body. 
We show the existence of ``no-shell'' solutions where
the field is nearly frozen in the whole interior of the body,
which does not necessarily correspond to 
the ``zero-shell" limit of thin-shell solutions.
In the no-shell case, under the condition $m_A r_c \gg 1$, 
the effective coupling of $\phi$ with matter
takes the same asymptotic form as that in the thin-shell case.
We study experimental bounds coming from the violation of 
equivalence principle as well as solar-system tests
for a number of models including $f(R)$ gravity
and find that the field is in either the thin-shell
or the no-shell regime under such constraints, 
depending on the shape of scalar-field potentials.
We also show that, for the consistency with local gravity 
constraints, the field at the center of the body needs to be 
extremely close to the value $\phi_A$
at the extremum of an effective potential induced by 
the matter coupling.

\end{abstract}

\date{\today}

\maketitle

\section{Introduction}

Recently there have been a lot of efforts to understand 
the origin of dark energy under the framework of
modified gravity theories (see Refs.~\cite{review} for reviews).
Presumably the simplest modified gravity model of dark energy 
is so-called $f(R)$ gravity in which $f$ is a function in 
terms of a Ricci scalar $R$ \cite{Capo}.
If we transform to the Einstein frame by a conformal transformation, 
$f(R)$ gravity is described by a scalar-field action with 
a potential that has gravitational origin \cite{Maeda}. 
In the Einstein frame, the scalar field couples to 
a non-relativistic matter with a coupling of the order 
of unity ($Q=-1/\sqrt{6}$) \cite{APT}. 
For such large coupling models the potential generally 
needs to be carefully designed to satisfy 
both cosmological and local gravity constraints.
See Refs.~\cite{AGPT,Hu,Star,Tsuji08} for the construction 
of viable $f(R)$ models.

Scalar-tensor models of dark energy \cite{stensor,TUMTY} 
also give rise to a similar coupling $Q$ in the Einstein frame, where 
the strength of $Q$ depends on the gravitational coupling 
with the scalar field in the Jordan frame \cite{coupled}.
Thus, in modified gravity models of dark energy, it is crucially 
important to appropriately study the compatibility of such 
couplings with local gravity experiments as well as with a
late-time acceleration of the Universe preceded by 
a standard matter era.

It is also known that there are many massless scalar fields 
such as a dilaton in superstring and supergravity 
theories \cite{Lidsey}.
In general, these fields couple to matter as well as 
gravity with strengths of the order of unity. 
The existence of such scalar fields 
gives rise to a strong violation of equivalence principle, 
thus posing a difficulty for the compatibility with 
local gravity experiments in the absence of field potentials.
One of the attempts to tackle this problem is the so-called
run-away dilaton scenario \cite{Damour} in which the field 
gradually decouples from matter in the large-field regime 
under an appropriate choice of the dilaton coupling.
It is then possible to apply the run-away dilaton scenario to 
dark energy provided that the dilaton has a monotonically 
decreasing potential toward larger $\phi$ \cite{Piazza}.

Another very interesting attempt to reconcile large coupling 
models with local gravity constraints is to use ``chameleon''
scalar fields whose masses depend on the environment they 
are in \cite{KW,KW2}. The chameleon mechanism naturally works for 
monotonically decreasing potentials that appear in the 
context of dark energy without adjusting the strength of 
the matter coupling $Q$ \cite{Brax}.
The chameleon mechanism has been applied to many models--
such as self-interacting scalar field models \cite{Gubser}, 
$f(R)$ gravity \cite{Hu,fR,Capo08}, scalar-tensor 
theories \cite{Clifton,TUMTY,Das}, curvature couplings to 
a scalar field \cite{Nava}, radions in braneworld models \cite{radion}, 
and theories whose matter coupling is much larger than the gravitational 
coupling \cite{Mota}.  
The presence of the matter coupling induces an extremum
of the field potential around which the field can stay.
If the density of the matter is sufficiently high as the interior
of a compact object, 
the field acquires a heavy mass about 
the potential minimum so that the violation of equivalence 
principle is suppressed even if $Q$ is of the order of unity.
Meanwhile the field has a lighter mass in a low-density 
cosmological environment
relevant to dark energy so that it can propagate freely.

In Refs.~\cite{KW,KW2} it was shown that local gravity 
constraints can be satisfied when a spherically symmetric 
body has a thin-shell. When the thin-shell is formed 
the chameleon field stays around the potential extremum 
at $\phi=\phi_A$ inside of the body 
in the region $0<r<r_1$, where $r_1$ is close to 
a radius $r_c$ of the body.
The field evolves in the thin-shell region characterized by 
$r_1<r<r_c$ due to the dominance of the matter coupling 
term $Q \rho_A$, where $\rho_A$ is a mean density of the body.
As long as the condition, $(r_c-r_1)/r_c \ll 1$, is satisfied,   
the effective coupling $Q_{\rm eff}$ outside of the body 
becomes much smaller than the order of $Q$ so that 
the models are consistent with local gravity experiments. 
See Refs.~\cite{exper} for recent experimental constraints 
on model parameters under the chameleon mechanism.

The analysis in Refs.~\cite{KW,KW2} assumes that the field 
is frozen ($\phi=\phi_A$) in the regime $0<r<r_1$ without 
explicitly taking into account the field mass $m_A$.  
In this paper we derive field solutions in three regimes
(a) $0<r<r_1$, (b) $r_1<r<r_c$, (c) $r>r_c$, 
and determine integration constants analytically by imposing 
appropriate boundary conditions at $r=0, r_1, r_c, \infty$, 
respectively. Under the condition that the field mass $m_B$ about the 
potential extremum far outside of the body is nearly massless, 
the field profile takes fairly simple forms as given 
in Eqs.~(\ref{ne1})-(\ref{ne3}).

We will show that standard thin-shell solutions found
in Refs.~\cite{KW,KW2} correspond to the limit $m_A r_c \to \infty$
with $(r_c-r_1)/r_c \ll 1$.
There exist ``no-shell'' solutions 
where the effect of the $Q \rho_A$ term does not become important 
in the whole interior of the body. 
These are wider class of solutions than the ``zero-shell" limit 
of thin-shell solutions as we show later. 
The effective coupling $Q_{\rm eff}$ outside
of the body in the no-shell case takes the same asymptotic form 
as that in the thin-shell case in the limit $m_A r_c \to \infty$.
Hence it is possible to satisfy local gravity constraints even if the body 
does not have a thin-shell.
We also study experimental bounds for concrete models and show 
that the field profile consistent with these bounds
corresponds to either thin-shell or no-shell solutions with $m_A r_c \gg 1$, 
depending upon the form of field potentials.

This paper is organized as follows.
In Sec.~\ref{mmodel} we provide a number of concrete models
which we will use in a later section.
These are basically motivated by the potentials 
that appear in the context of dark energy.
In Sec.~\ref{chame} we review the chameleon mechanism and 
derive new solutions by properly considering the field profile 
in the region $0<r<r_1$.
In Sec.~\ref{nosh} we obtain three different solutions:
(i) thick-shells, (ii) thin-shells and (iii) no-shells, by taking some 
limit for the solutions derived in Sec.~\ref{chame}.
In Sec.~\ref{concrete} we apply our formula to
a number of field potentials and clarify the regime of the field profile
when local gravity constraints are satisfied.
Sec.~\ref{conclude} is devoted to conclusions.

\section{Models}
\label{mmodel}

We consider a scenario in which a scalar field $\phi$ with 
potential $V(\phi)$ couples to a non-relativistic matter with 
a Lagrangian density ${\cal L}_m$. The action we study is given by 
\begin{eqnarray}
\label{action}
S &=& \int {\rm d}^4 x\sqrt{-g} 
\left[ \frac{M_{\rm pl}^2}{2} R
-\frac12 (\nabla \phi)^2-V(\phi) \right]
-\int {\rm d}^4x\,{\cal L}_m 
(\Psi_m^{(i)},g_{\mu \nu}^{(i)})\,,
\end{eqnarray}
where $g$ is a determinant of the metric $g_{\mu \nu}$, 
$M_{\rm pl}=1/\sqrt{8 \pi G}$ is a reduced Planck mass
($G$ is gravitational constant), 
$R$ is a Ricci scalar, and $\Psi_m^{(i)}$ are matter fields
that couple to a metric $g_{\mu \nu}^{(i)}$ related with 
the Einstein frame metric $g_{\mu \nu}$ via
\begin{eqnarray}
\label{gmunu1}
g_{\mu \nu}^{(i)}=e^{2 Q_i \phi} g_{\mu \nu}\,.
\end{eqnarray}
Here $Q_i$ are the strength of couplings 
for each matter field.
In the following we shall use the unit $M_{\rm pl}=1$, but 
we restore $M_{\rm pl}$ or $G$ when it is needed.

There are theories that give rise to a constant coupling 
$Q$ for each matter field.
Let us consider the following scalar-tensor action \cite{TUMTY}
\begin{eqnarray}
\label{actionsca}
\tilde{S} &=& \int {\rm d}^4 x\sqrt{-\tilde{g}} 
\biggl[ \frac12 F(\phi) \tilde{R}
-\frac12 (1-6Q^2)F(\phi) (\tilde{\nabla} \phi)^2
-U(\phi) \biggr] -\int {\rm d}^4x\,{\cal L}_m 
(\Psi_m, \tilde{g}_{\mu \nu})\,,
\end{eqnarray}
where a tilde represents quantities in the Jordan frame and 
\begin{eqnarray}
F(\phi)=e^{-2Q \phi}\,.
\end{eqnarray}
Under a conformal transformation,  
\begin{eqnarray}
\label{gmunu2}
g_{\mu \nu}=F(\phi) \tilde{g}_{\mu \nu}\,,
\end{eqnarray}
we get the Einstein frame action (\ref{action}) with 
the field potential
\begin{eqnarray}
V(\phi)=U(\phi)/F(\phi)^2\,.
\end{eqnarray}
It is worth mentioning that the action (\ref{actionsca})
is equivalent to the following Brans-Dicke theory \cite{Brans}
with a potential $U(\phi)$:
\begin{eqnarray}
\label{action0}
\tilde{S} = \int {\rm d}^4 x\sqrt{-\tilde{g}} 
\left[ \frac12 \chi \tilde{R}
-\frac{\omega_{\rm BD}}{2\chi} (\tilde{\nabla} \chi)^2 
-U(\phi(\chi)) \right] 
-\int {\rm d}^4x\,{\cal L}_m 
(\Psi_m, \tilde{g}_{\mu \nu})\,,
\end{eqnarray}
where $\chi=F(\phi)=e^{-2Q \phi}$ and
$\omega_{\rm BD}$ is a Brans-Dicke parameter that is 
related with the coupling $Q$
via the relation $3+2\omega_{\rm BD}=1/(2Q^2)$ \cite{TUMTY}.
The $f(R)$ gravity corresponds to the coupling $Q=-1/\sqrt{6}$, 
i.e., $\omega_{\rm BD}=0$ \cite{Chiba03}.

If the field potential is absent, solar-system experiments
constrain the Brans-Dicke parameter to be 
$\omega_{\rm BD}>4.0 \times 10^4$ \cite{lgccon}.
This translates into the bound: $|Q|<2.5 \times 10^{-3}$.
When $|Q|$ is of the order of unity, as in the case of $f(R)$ gravity, 
the presence of the potential is essentially important for the 
consistency with local gravity experiments.
If the scalar field is responsible for dark energy, 
it is natural to design the potential $V(\phi)$ so that the
field runs away toward larger $|\phi|$ with a tracking 
behavior at late times.
An example of this type is the inverse power-law 
potential \cite{Ratra}
\begin{eqnarray}
\label{po1}
V(\phi)=M^{4+n} \phi^{-n}\,,
\end{eqnarray}
where $M$ has a unit of mass and $n$ is a constant.
The scalar field is almost trapped at the extremum of 
an effective potential induced by the presence of 
the matter coupling $Q$ with $|\phi|$ much smaller 
than $M_{\rm pl}$ (as we will see later).
As long as the field stays at the extremum in the region of high density, 
it is possible to satisfy local gravity constraints through 
the chameleon mechanism.

In the context of $f(R)$ gravity, Hu and Sawicki \cite{Hu} and
Starobinsky \cite{Star} proposed models that can be 
consistent with cosmological and local gravity constraints.
In the region where $\tilde{R}$ is much larger than a critical 
value $\tilde{R}_c$ ($R_c$ is of the order of the present 
cosmological Ricci scalar), these models correspond to 
the Jordan frame action 
\begin{eqnarray}
\label{fRJo}
\tilde{S} = \int {\rm d}^4 x\sqrt{-\tilde{g}} 
\,\frac12 f(\tilde{R}) -\int {\rm d}^4x\,{\cal L}_m 
(\Psi_m,\tilde{g}_{\mu \nu})\,,
\end{eqnarray}
with
\begin{eqnarray}
\label{fR}
f(\tilde{R}) = \tilde{R}-\mu \tilde{R}_c [1-
(\tilde{R}/\tilde{R}_c)^{-2n}]\,,
\end{eqnarray}
where $\mu$ and $n$ are positive constants \cite{Tsuji08}.
Comparing the action (\ref{fRJo}) with (\ref{actionsca}) 
we find that the potential in $f(\tilde{R})$ gravity ($Q=-1/\sqrt{6}$) 
corresponds to $U=(F\tilde{R}-f)/2$ with a dynamical scalar field
$\phi \equiv (\sqrt{6}/2)\,{\rm ln}\,F$, where 
$F$ is related with $f$ via the relation 
$F=\partial f/\partial \tilde{R}$ \cite{TUMTY}.
The field potential for the model (\ref{fR}) is given by 
\begin{eqnarray}
\label{po2}
U(\phi)=\frac{\mu \tilde{R}_c}{2}
\left[ 1-\frac{2n+1}{(2n\mu)^{2n/(2n+1)}} 
(1-e^{2\phi/\sqrt{6}})^{\frac{2n}{2n+1}}\right]\,.
\end{eqnarray}

For the action (\ref{actionsca}) with arbitrary 
couplings $Q$, one can also construct viable models 
by generalizing the potential (\ref{po2}) in $f(R)$ gravity.
An explicit example of this type of potential
is given by \cite{TUMTY}
\begin{eqnarray}
\label{po}
U(\phi)=V_0 \left[ 1-C (1-e^{-2Q\phi})^p \right]\,, 
\end{eqnarray}
where $V_0>0,~C>0,~0<p<1$.
The model (\ref{fR}) in $f(R)$ gravity corresponds to 
$p=2n/(2n+1)$.
The scalar field mass gets larger as $p$ becomes closer to 1.
Since $|\phi|$ is much smaller than $1$
in the region of high density, the potential 
$V(\phi)=e^{4Q\phi}U(\phi)$ in the Einstein frame
is almost identical to the  potential $U(\phi)$ in the Jordan frame
for $|Q| \lesssim 1$.

If the action (\ref{action}) does not originate from 
the scalar-tensor action, it can happen that the strength of the 
coupling $Q$ is not the same between different matter species. 
For example, in coupled dark energy scenario in Ref.~\cite{coupled},
it is assumed that a quintessence field couples to dark matter
but not to baryons.

\section{Chameleon mechanism}
\label{chame}

In this section we revisit the chameleon mechanism paying particular
attention to the scalar-field mass inside of a spherically symmetric
body. The contribution of metric perturbations to the scalar-field 
equation is neglected as we only consider a weak gravitational background. 
As a matter source we take into account a non-relativistic fluid 
whose pressure is negligible relative to its energy density.

Varying the Einstein frame action (\ref{action}) with respect to the 
field $\phi$, we get 
\begin{equation}
\label{ori}
\square \phi-V_{,\phi}
=-\sum_i Q_i e^{4Q_i \phi} g^{\mu \nu}_{(i)}
T_{\mu \nu}^{(i)}\,,
\end{equation}
where $T_{\mu \nu}^{(i)}=(2/\sqrt{-g^{(i)}})
\delta {\cal L}_m/\delta g^{\mu \nu}_i$ is the energy 
momentum tensor of the $i$-th matter.
The trace of the $i$-th matter is given by $T^{(i)} \equiv
g^{\mu \nu}_{(i)}T_{\mu \nu}^{(i)}=-\tilde{\rho}_i$
for a non-relativistic fluid, where $\tilde{\rho}_i$ is 
an energy density.

{}From the comparison of Eq.~(\ref{gmunu1}) 
with Eq.~(\ref{gmunu2}), it is clear that $\tilde{\rho}_i$
has a meaning of the energy density in the Jordan frame
if the action (\ref{action}) originates from (\ref{actionsca}) 
under the conformal transformation.
The energy density in the Einstein frame corresponds 
to $\rho_i^{({\rm E})}=\tilde{\rho}_i\, e^{4Q_i \phi}$, 
but this does not satisfy the usual continuity equation.
It is more convenient to introduce the quantity $\rho_i=
\tilde{\rho}_i\, e^{3Q_i \phi}$, which is conserved 
in the Einstein frame\footnote{In the Friedmann-Robertson-Walker
cosmological background this means that 
$\rho_i$ satisfies the equation $\dot{\rho}_i+3H \rho_i=0$ 
($H$ is a Hubble parameter), while the equation for $\rho_i^{(\rm E)}$ is 
$\dot{\rho_i}^{({\rm E})}+3H\rho_i^{({\rm E})}
=Q_i \dot{\phi}\rho_i^{({\rm E})}$ \cite{Waterhouse}.}.
Then Eq.~(\ref{ori}) reduces to 
\begin{eqnarray}
\label{ori2}
\square \phi=V_{,\phi}
+\sum_i Q_i \rho_i e^{Q_i \phi}\,.
\end{eqnarray}

In a spherically symmetric background Eq.~(\ref{ori2}) yields
\begin{eqnarray}
\label{ori3}
\frac{\rd^2 \phi}{\rd r^2}+\frac{2}{r} \frac{\rd \phi}{\rd r}
=\frac{\rd V_{\rm eff}}{\rd \phi}\,,
\end{eqnarray}
where $r$ is a distance from the center of symmetry and 
the effective potential $V_{\rm eff}$ is defined by 
\begin{eqnarray}
V_{\rm eff} (\phi) \equiv V(\phi)+\sum_i \rho_i e^{Q_i \phi}\,.
\end{eqnarray}
In the following, unless otherwise stated, we shall consider the case 
in which couplings $Q_i$ are the same for each matter component, 
i.e., $Q_i=Q$ and $\rho_i=\rho$.

When $Q>0$ the potential has a minimum for the models 
with $V_{,\phi}<0$. For example, the potential (\ref{po1})
gives rise to a minimum at 
\begin{eqnarray}
\phi_M \simeq \left[ \frac{n}{Q} 
\frac{M_{\rm pl}^4}{\rho} \left( \frac{M}{M_{\rm pl}}
\right)^{4+n} \right]^{1/(n+1)}M_{\rm pl}\,,
\end{eqnarray}
where we recovered the reduced Planck mass $M_{\rm pl}$.
The condition, $\phi_M \ll M_{\rm pl}$, needs to be satisfied 
for the consistency with local gravity constraints \cite{KW,KW2}.

When $Q<0$, as in $f(R)$ gravity, the potential has a minimum 
for the models with $V_{,\phi}>0$.
In fact the potential given in Eq.~(\ref{po}), i.e.,
$V(\phi)=V_0 e^{4Q \phi} \left[ 1-C (1-e^{-2Q \phi})^p \right]$,
satisfies the condition $V_{,\phi}>0$ for $Q<0$.
Then the effective potential $V_{\rm eff}$ has a minimum at 
\begin{eqnarray}
\phi_M \simeq 
\frac{1}{2Q} \left( \frac{2pC V_0}{\rho} \right)^{1/(1-p)}
M_{\rm pl}\,,
\end{eqnarray}
which exists in the region $\phi_M<0$.
Note that the order of $V_0$ is not much different from 
the present cosmological density 
$\rho_0 \simeq 10^{-29}$\,g/cm$^3$
if this potential is responsible for the accelerated 
expansion today.
Hence in the region $\rho \gg V_0$ the condition 
$|\phi_M| \ll M_{\rm pl}$ is well satisfied \cite{TUMTY}.

In the following we assume that the spherically symmetric 
body has a homogeneous density $\rho=\rho_A$
and that the density is homogeneous with a value $\rho=\rho_B$ 
outside of the body. 
The mass of this body is given by $M_c=(4\pi/3) \rho_A r_c^3$, 
where $r_c$ is a radius of the body.
The effective potential $V_{\rm eff}$ has minima at field 
values $\phi_A$ and $\phi_B$ characterized 
by the conditions
\begin{eqnarray}
& & V_{,\phi} (\phi_A)+Q \rho_A e^{Q \phi_A}=0\,,\\
& &V_{,\phi} (\phi_B)+Q\rho_B e^{Q \phi_B}=0\,.
\end{eqnarray}
The former corresponds to the region with a high density (interior of the body)
that gives rise to a heavy mass squared 
$m_A^2 \equiv \frac{\rd^2 V_{{\rm eff}}}{\rd\phi^2}(\phi_A)$,
whereas the latter to the lower density region  (exterior of the body) 
with a lighter mass squared 
$m_B^2 \equiv \frac{\rd^2 V_{{\rm eff}}}{\rd\phi^2}(\phi_B)$.

We impose the following boundary conditions:
\begin{eqnarray}
\frac{{\rm d}\phi}{{\rm d} r}(r=0)=0\,,\quad
\phi (r \to \infty)=\phi_B\,.
\end{eqnarray}
Equation (\ref{ori3}) shows that we need to consider the 
potential $(-V_{\rm eff})$ in order to find the ``dynamics''
of $\phi$ with respect to $r$. This means that the effective 
potential $(-V_{\rm eff})$ has a maximum at $\phi=\phi_A$.
The field $\phi$ is at rest at $r=0$ and begins to roll
down the potential when the matter-coupling term 
$Q \rho_A e^{Q \phi}$ becomes important at a radius $r_1$.

If the field value at $r=0$ is close to $\phi_A$, the field stays
around $\phi_A$ in the region $0<r<r_1$.
The body has a thin-shell if $r_1$ is close to the radius 
$r_c$ of the body. Whether $r_1$ can be close to $r_c$ or not 
also depends on the negative mass squared $(-m_A^2)$ 
at the potential maximum.
If $m_A$ is much larger than $1/r_c$, it is likely that we need
to determine the boundary condition of $\phi$ at $r=0$
very close to $\phi_A$ in order to obtain thin-shell solutions.

In the region $0 < r<r_1$, the r.h.s. of Eq.~(\ref{ori3}) can be approximated 
as $\rd V_{\rm eff}/\rd \phi \simeq m_A^2 (\phi-\phi_A)$ around 
$\phi=\phi_A$. Then the solution of Eq.~(\ref{ori3}) is given by 
$\phi(r)=\phi_A+A e^{-m_A r}/r+B e^{m_A r}/r$, where 
$A$ and $B$ are integration constants.
To avoid the divergence of $\phi$ at $r=0$, we require that 
$B=-A$. Then the solution is  
\begin{eqnarray}
\label{sol1}
\phi(r)=\phi_A+\frac{A (e^{-m_A r}-e^{m_A r})}{r}~~~~~~
(0 < r <r_1).
\end{eqnarray}
This automatically satisfies the boundary condition: 
$\frac{\rd \phi}{\rd r} (r=0)=0$.

In the region $r_1<r<r_c$ the field $|\phi(r)|$ evolves toward larger values
with the increase of $r$.
Since $|V_{,\phi}| \ll |Q \rho_A e^{Q \phi}|$ in this regime one has
$\rd V_{\rm eff}/\rd \phi \simeq Q \rho_A$ in Eq.~(\ref{ori3}), where 
we used the condition $Q\phi \ll 1$.
Hence we obtain the following solution 
\begin{eqnarray}
\label{sol2}
\phi(r)=\frac16 Q \rho_A r^2 -\frac{C}{r}+D
~~~~~~(r_1 < r <r_c),
\end{eqnarray}
where $C$ and $D$ are constants.

In the region outside of the body the field $|\phi|$ climbs up 
the potential hill toward larger values.
The kinetic energy of the field dominates over the
potential energy, which means that the r.h.s. of
Eq.~(\ref{ori3}) can be neglected relative to each component of 
the l.h.s. of it. Taking into account the mass term
$\rd V_{\rm eff}/\rd \phi \simeq m_B^2 (\phi-\phi_B)$
on the  r.h.s. of Eq.~(\ref{ori3}), we obtain the solution 
$\phi(r)=\phi_B+E e^{-m_B (r-r_c)}/r+F e^{m_B (r-r_c)}/r$
with integration constants $E$ and $F$.
Demanding the boundary condition, $\phi (r \to \infty)=\phi_B$, 
one has $F=0$ and hence the solution is given by 
\begin{eqnarray}
\label{sol3}
\phi(r)=\phi_B+E \frac{e^{-m_B (r-r_c)}}{r}~~~~~~(r > r_c).
\end{eqnarray}

We match three solutions (\ref{sol1}), (\ref{sol2}) and (\ref{sol3})
by imposing continuous conditions for $\phi$ and $\rd \phi/\rd r$
at $r=r_1$ and $r=r_c$.
Then four coefficients $A$, $C$, $D$ and $E$ 
are determined accordingly:
\begin{eqnarray}
\label{re1}
C &=& \frac{s_1 s_2 [(\phi_B-\phi_A)+(r_1^2-r_c^2)Q \rho_A/6]
+[ s_2 r_1^2 (e^{-m_A r_1}-e^{m_A r_1})-s_1 r_c^2] Q\rho_A/3}
{m_A (e^{-m_A r_1}+e^{m_A r_1})s_2-m_B s_1}\,,\\
\label{re2}
A &=& -\frac{1}{s_1} (C+Q \rho_A r_1^3/3)\,,\\
\label{re3}
E &=& -\frac{1}{s_2}(C+Q \rho_A r_c^3/3)\,, \\
\label{re4}
D &=& \phi_B -Q \rho_A r_c^2/6+
\frac{1}{r_c}(C+E)\,,
\end{eqnarray}
where 
\begin{eqnarray}
s_1 &\equiv& m_A r_1 (e^{-m_A r_1}+e^{m_A r_1})+
e^{-m_A r_1}-e^{m_A r_1}\,,\\
s_2 &\equiv& 1+m_B r_c\,.
\end{eqnarray}

Especially when the conditions, $m_B r_c \ll 1$ and $m_A \gg m_B$,
are satisfied so that the contribution of the $m_B$-dependent terms 
are negligible, we have
\begin{eqnarray}
\label{me1}
A &=& -\frac{1}{m_A (e^{-m_A r_1}+e^{m_A r_1})}
\left[ \phi_B-\phi_A+\frac12 Q \rho_A (r_1^2-r_c^2) \right]\,,\\
\label{me2}
C &=& r_1 \left[ 1+\frac{e^{-m_A r_1}-e^{m_Ar_1}}
{m_A r_1 (e^{-m_A r_1}+e^{m_A r_1})} \right]
\left[ \phi_B-\phi_A+\frac12 Q \rho_A (r_1^2-r_c^2) \right]
-\frac13 Q \rho_A r_1^3
\,,\\
\label{me3}
D &=& \phi_B-\frac12 Q \rho_A r_c^2\,, \\
\label{me4}
E &=& -r_1 (\phi_B -\phi_A)-\frac16 Q \rho_A r_c^3
\left(2+\frac{r_1}{r_c} \right) \left(1-\frac{r_1}{r_c} \right)^2
-\frac{e^{-m_A r_1}-e^{m_Ar_1}}
{m_A (e^{-m_A r_1}+e^{m_A r_1})} 
\left[ \phi_B-\phi_A+\frac12 Q \rho_A (r_1^2-r_c^2) \right].
\end{eqnarray}
This leads to the following solution 
\begin{eqnarray}
\label{ne1}
\phi(r) &=& \phi_A-\frac{1}{m_A (e^{-m_A r_1}+e^{m_A r_1})}
\left[ \phi_B-\phi_A+\frac12 Q \rho_A (r_1^2-r_c^2) \right]
\frac{e^{-m_A r}-e^{m_A r}}{r}~~~~~~~~(0<r<r_1),\\
\label{ne2}
\phi(r) &=& \phi_B+\frac16 Q \rho_A (r^2-3r_c^2)
+\frac{Q \rho_A r_1^3}{3r} \nonumber \\
& & -\left[ 1+\frac{e^{-m_A r_1}-e^{m_Ar_1}}
{m_A r_1 (e^{-m_A r_1}+e^{m_A r_1})} \right]
\left[ \phi_B-\phi_A+\frac12 Q \rho_A (r_1^2-r_c^2) \right]
\frac{r_1}{r}~~~~~~~~(r_1<r<r_c),\\
\label{ne3}
\phi(r) &=& \phi_B - \Biggl[ 
r_1 (\phi_B -\phi_A)+\frac16 Q \rho_A r_c^3
\left(2+\frac{r_1}{r_c} \right) \left(1-\frac{r_1}{r_c} 
\right)^2 \nonumber \\
& &~~~~~~~~ +\frac{e^{-m_A r_1}-e^{m_Ar_1}}
{m_A  (e^{-m_A r_1}+e^{m_A r_1})} 
\left\{ \phi_B-\phi_A+\frac12 Q \rho_A (r_1^2-r_c^2)
\right\} \Biggr] \frac{e^{-m_B (r-r_c)}}{r}
~~~~~~~~(r>r_c)\,.
\end{eqnarray}
While a similar matching procedure given above was considered 
in Ref.~\cite{Waterhouse}, the analytic expression 
(\ref{ne1})-(\ref{ne3}) has not been derived so far.

The radius $r_1$ is determined by the following condition
\begin{eqnarray}
m_A^2 \left[ \phi(r_1) -\phi_A \right]=Q \rho_A,
\end{eqnarray}
which translates into
\begin{eqnarray}
\label{r1con}
 \phi_B -\phi_A +\frac12 Q \rho_A 
(r_1^2-r_c^2) =\frac{6Q \Phi_c}{(m_A r_c)^2}
\frac{m_A r_1 (e^{m_A r_1}+e^{-m_A r_1})}
{e^{m_A r_1}-e^{-m_A r_1}}\,,
\end{eqnarray}
where $\Phi_c=M_c/(8\pi r_c)=\rho_A r_c^2/6$
is a gravitational potential at the surface of the body.
Under this relation the field profile (\ref{ne3})
outside of the body can be written as 
\begin{eqnarray}
\label{fieldout}
 \phi (r)=\phi_B -2Q\Phi_c r_c
 \left[ 1 -\frac{r_1^3}{r_c^3} +3\frac{r_1}{r_c}
 \frac{1}{(m_A r_c)^2} \left\{
 \frac{m_A r_1 (e^{m_A r_1}+e^{-m_A r_1})}
 {e^{m_A r_1}-e^{-m_A r_1}}-1 \right\} \right]
 \frac{e^{-m_B (r-r_c)}}{r}
 ~~~~~(r>r_c)\,.
\end{eqnarray}
%

\section{Thick-shell, thin-shell, and no-shell solutions}
\label{nosh}

In this section we derive three different types of solutions: 
(i) thick-shells, (ii) thin-shells, and (iii) no-shells
by taking appropriate limits for the results 
derived in the previous section.

\subsection{Thick-shell solutions ($r_1 \to 0$)}

If the field value at $r=0$ is away from $\phi_A$, it happens 
that the field rolls down the potential for $r \ge 0$.
This is the thick-shell regime in which the field inside 
of the body ($0<r<r_c$) changes significantly.

More precisely the following relation holds in the 
thick-shell case:
\begin{eqnarray}
\label{thiccon}
m_A^2 \left[ \phi (0)-\phi_A \right]>Q \rho_A\,.
\end{eqnarray}
The field value at $r=0$ is given by $\phi(0)=
\phi_B-(1/2)Q \rho_A r_c^2$ by taking the limit 
$r_1 \to 0$ in Eq.~(\ref{ne1}).
Then the condition (\ref{thiccon}) can be expressed as
\begin{eqnarray}
\label{epre}
\epsilon_{\rm th}
>\frac12+\frac{1}{(m_A r_c)^2}\,,
\end{eqnarray}
where $\epsilon_{\rm th}$ is the so-called thin-shell 
parameter \cite{KW,KW2} defined by 
\begin{eqnarray}
\epsilon_{\rm th} \equiv 
\frac{\phi_B-\phi_A}{6Q \Phi_c}\,.
\end{eqnarray}
Equation (\ref{epre}) shows that, in the thick-shell case, 
$\epsilon_{\rm th} $ is larger than the order of unity.

Taking the limit $r_1 \to 0$ in Eqs.~(\ref{ne2}) 
and (\ref{ne3}), we get 
\begin{eqnarray}
& &\phi(r)=\phi_B+Q \Phi_c \left( \frac{r^2}{r_c^2}
-3 \right)~~~~~~
(0<r<r_c),\\
\label{thickQ}
& & \phi(r)=\phi_B- 2Q \frac{GM_c}{r} 
e^{-m_B(r-r_c)}~~~~~~(r>r_c)\,.
\end{eqnarray}
{}From Eq.~(\ref{thickQ}) the effective coupling outside of the body 
is of the order of $Q$, which means that local gravity constraints are not 
satisfied unless $|Q|$ is very much smaller than unity.

\subsection{Thin-shell solutions ($(r_c-r_1) \ll r_c$)}

{}From Eq.~(\ref{fieldout}) we find that it is possible to make 
the effective coupling small if $r_1$ is close to $r_c$.
This corresponds to the thin-shell regime in which 
the field is stuck in the region $0<r<r_1$ with 
$(r_c-r_1)/r_c \ll 1$.
Equation (\ref{fieldout}) shows that the mass term $m_A$
also affects the strength of the effective coupling.
In the following we shall discuss the cases of large and 
small mass limits separately.

\subsubsection{The massive case $(m_A r_c \gg 1)$}

As we see below, thin-shell solutions originally derived 
in Refs.~\cite{KW,KW2} can be recovered by taking the limit 
\begin{eqnarray}
m_A r_1 \gg 1\,,
\end{eqnarray}
together with the thin-shell condition given by 
\begin{eqnarray}
\Delta r_c \equiv r_c-r_1 \ll r_c\,.
\end{eqnarray}
Expanding Eq.~(\ref{r1con}) in terms of small parameters
$\Delta r_c/r_c$ and $1/m_A r_c$, we obtain 
\begin{eqnarray}
\label{epthin}
\epsilon_{\rm th} \simeq 
\frac{\Delta r_c}{r_c}+\frac{1}{m_A r_c}\,.
\end{eqnarray}
When $m_A r_c \gg (\Delta r_c/r_c)^{-1}$, this recovers 
the relation $\epsilon_{\rm th} \simeq \Delta r_c/r_c$
derived in Refs.~\cite{KW,KW2}.

{}From Eq.~(\ref{fieldout}) the field profile outside of 
the body is given by 
\begin{eqnarray}
\label{ne3d}
\phi(r) \simeq \phi_B-6 Q \Phi_c r_c
\left[ \frac{\Delta r_c}{r_c}+\frac{1}{m_A r_c}
-\left( \frac{\Delta r_c}{r_c} \right)^2
-\frac{1}{(m_A r_c)^2} 
-\frac{2}{m_Ar_c}\frac{\Delta r_c}{r_c}
\right]
\frac{e^{-m_B (r-r_c)}}{r}\,.
\end{eqnarray}
Neglecting second-order terms with respect to 
$\Delta r_c/r_c$ and $1/m_A r_c$, we get 
\begin{eqnarray}
\label{phir}
\phi(r) \simeq \phi_B -2Q_{\rm eff} \frac{GM_c}{r}
e^{-m_B (r-r_c)}\,,
\end{eqnarray}
where $Q_{\rm eff}$ is the effective coupling given by 
\begin{eqnarray}
\label{Qeff0}
Q_{\rm eff} \simeq 3Q 
\left( \frac{\Delta r_c}{r_c}+\frac{1}{m_A r_c}
\right)=3Q \epsilon_{\rm th}\,,
\end{eqnarray}
where we used the relation (\ref{epthin}).
As long as both $\Delta r_c/r_c$ and $1/(m_A r_c)$
are much smaller than unity so that $\epsilon_{\rm th} \ll 1$, 
it is possible to satisfy local gravity constraints.

On using Eqs.~(\ref{ne1}), (\ref{ne2}) and (\ref{ne3d})
together with Eq.~(\ref{r1con}), we find that 
the field value at $r=0$, $r=r_1$, $r=r_c$ 
are given, respectively, by 
\begin{eqnarray}
\label{phi0}
& &\phi(0) \simeq \phi_A +\frac{12 Q\Phi_c}{m_A r_c e^{m_A r_c}}\,, \\
\label{phi1}
& &\phi(r_1) \simeq  \phi_A +\frac{6Q \Phi_c}{(m_A r_c)^2}\,,\\
\label{phic}
& &\phi(r_c) \simeq \phi_A+6Q\Phi_c \left[ 
\frac{1}{(m_A r_c)^2}+
\frac{1}{m_A r_c} \frac{\Delta r_c}{r_c}+
\frac12 \left( \frac{\Delta r_c}{r_c} \right)^2 \right]\,.
\end{eqnarray}
Due to the presence of the $e^{m_A r_c}$ term in Eq.~(\ref{phi0}),
$\phi (0)$ needs to be very close to $\phi_A$.
Under the condition $m_A r_c \gg 1$, the field can rapidly roll down 
the potential unless it is very close to the potential maximum 
at $\phi=\phi_A$.
Note also that when $Q<0$ (including $f(R)$ gravity) 
we have $\phi(0)<\phi_A~(<0)$ from (\ref{phi0}).
Hence the boundary condition at $r=0$ that corresponds 
to thin-shell solutions with $m_A r_c \gg 1$ 
is chosen so that the field does not 
reach the curvature singularity at $\phi=0$ \cite{Frolov}.

{}From Eqs.~(\ref{phi0})-(\ref{phic}) it is clear that 
$|\phi|$ gradually increases for larger $r$.
The field stays around $\phi=\phi_A$ within the body 
and it changes to the value $\phi_B$ for $r>r_c$.

\subsubsection{The light mass case $(m_A r_c \ll 1)$}

Let us next consider the light mass case satisfying the condition 
\begin{eqnarray}
m_A r_1 \ll 1\,,
\end{eqnarray}
together with the thin-shell condition $\Delta r_c \ll r_c$.
Note that this case is not equivalent to the thick-shell case in which 
the limit $r_1 \to 0$ is taken in Eqs.~(\ref{ne1})-(\ref{ne3}).
Since the radius $r_1$ is close to $r_c$, the $r_1$-dependent term
such as $Q \rho_A r_1^3/(3r)$ in Eq.~(\ref{ne2}) is non-vanishing.

{}From Eq.~(\ref{r1con}) we get the following relation 
\begin{eqnarray}
\label{ep2}
\epsilon_{\rm th} \simeq \frac{\Delta r_c}{r_c}+
\frac{1}{(m_A r_c)^2} \simeq 
\frac{1}{(m_A r_c)^2}\,,
\end{eqnarray}
which gives the relation $\epsilon_{\rm th} \gg 1$.
Even if the body has a thin-shell ($\Delta r_c/r_c \ll 1$),
the parameter $\epsilon_{\rm th}$ is much larger than unity
under the condition $m_A r_c \ll 1$.
The field profile (\ref{fieldout}) in the region $r>r_c$ reduces to 
\begin{eqnarray}
\phi (r) \simeq \phi_B 
-2Q \left[ 1-\frac{1}{15} (m_A r_c)^2 \right]  
\frac{GM_c}{r} e^{-m_B (r-r_c)}\,.
\end{eqnarray}
This means that the coupling is of the order of $Q$
as in the thick-shell case.
Hence it is not possible to be compatible with 
local gravity constraints for $|Q|={\cal O}(1)$.
The effective coupling $Q_{\rm eff}$ is of the order of $Q$ even if $r_1$ exists
in the intermediate regime ($0<r_1<r_c$) away from $r_c$.

The field value at $r=r_c$ is estimated as
\begin{eqnarray}
\phi(r_c) \simeq \phi_B-2Q \Phi_c \simeq \phi_B\,,
\end{eqnarray}
where we used the relation 
$\phi_B \simeq 6Q \Phi_c/(m_A r_c)^2$
that comes from Eq.~(\ref{ep2}) under the condition 
$|\phi_B| \gg |\phi_A|$.
Unlike thin-shell solutions with a massive limit ($m_A r_c \gg 1$), 
the field stays around $\phi \simeq \phi_B$
during the transition from the inside to the outside of the body.
This property is different from the thin-shell case with the
massive limit in which the field changes from $\phi_A$ ($0<r<r_c$) 
to $\phi_B$ ($r>r_c$).

\subsection{No-shell solutions ($r_1=r_c$)}

Let us next consider the case in which the solution is 
described by Eq.~(\ref{ne1}) in the whole internal region of the body.
In this case Eq.~(\ref{ne1}) is joined to Eq.~(\ref{ne3})
at the surface of the body ($r=r_c$).
Setting $r_1=r_c$ in Eqs.~(\ref{ne1}) and (\ref{ne3}), we get 
the following field profile:
\begin{eqnarray}
\label{nosh1}
& &\phi(r) = \phi_A-\frac{\phi_B-\phi_A}{m_A (e^{-m_A r_c}+e^{m_A r_c})}
\frac{e^{-m_A r}-e^{m_A r}}{r}~~~~~~(0<r<r_c)\,,\\
\label{nosh2}
& &\phi (r) =\phi_B -\left[ r_c+\frac{e^{-m_A r_c}-e^{m_A r_c}}
{m_A (e^{-m_A r_c}+e^{m_A r_c})} \right] (\phi_B-\phi_A)
\frac{e^{-m_B (r-r_c)}}{r}~~~~~~(r>r_c)\,,
\end{eqnarray}
which agrees with the result in Ref.~\cite{Waterhouse}.
We require that the following condition is satisfied:
\begin{eqnarray}
m_A^2 \left[ \phi (r_c)-\phi_A \right]\leq Q \rho_A\,,
\end{eqnarray}
which is equivalent to 
\begin{eqnarray}
\label{epcon}
\epsilon_{\rm th}\leq \frac{e^{m_A r_c}+e^{-m_A r_c}}
{m_A r_c (e^{m_A r_c}-e^{-m_A r_c})}\,.
\end{eqnarray}
Notice the essential difference from the ``zero-shell" limit of 
the thin-shell case where the equality holds in Eq.~(\ref{epcon}). 

\subsubsection{The massive case $(m_A r_c \gg 1)$}

If the field $\phi$ is massive such that $m_A r_c \gg 1$, 
the solution outside of the body is 
\begin{eqnarray}
\phi (r) \simeq \phi_B -6Q \frac{GM_c}{r} \epsilon_{\rm th}
\left( 1-\frac{1}{m_A r_c} \right) e^{-m_B (r-r_c)}\,.
\end{eqnarray}
This shows that the effective coupling is given by 
$Q_{\rm eff} \simeq 3Q \epsilon_{\rm th}$, 
which is the same as Eq.~(\ref{Qeff0}) in the thin-shell case 
with the massive limit ($m_A r_c \gg 1$). 
It is possible to satisfy local gravity constraints 
provided that $\epsilon_{\rm th} \ll 1$.
The field values at $r=0$ and $r=r_c$ are
\begin{eqnarray}
\label{phi0no}
& &\phi (0) \simeq \phi_A+\frac{2 (\phi_B-\phi_A)}{e^{m_A r_c}}\,,\\
& &\phi(r_c) \simeq \phi_A+\frac{\phi_B-\phi_A}{m_A r_c}\,.
\end{eqnarray}
Hence $\phi(0)$ needs to be very close to $\phi_A$.

Equation (\ref{epcon}) gives the following constraint
\begin{eqnarray}
\epsilon_{\rm th}\leq \frac{1}{m_A r_c}\,.
\end{eqnarray}
{}From Eq.~(\ref{epthin}) we find that the opposite inequality, 
$\epsilon_{\rm th}> 1/(m_A r_c)$, holds for the thin-shell case
in the massive limit ($m_A r_c \gg 1$).

\subsubsection{The light mass case $(m_A r_c \ll 1)$}

When the field is almost massless such that $m_A r_c \ll 1$, 
the solution in the region $r>r_c$ is
\begin{eqnarray}
\label{massless}
\phi(r) \simeq \phi_B -2Q \frac{GM_c}{r} 
\epsilon_{\rm th} (m_A r_c)^2 
e^{-m_B (r-r_c)}\,.
\end{eqnarray}
The field stays around the value $\phi_B$
both inside and outside of the body.

{}From Eq.~(\ref{epcon}) we get 
\begin{eqnarray}
\epsilon_{\rm th}\leq \frac{1}{(m_A r_c)^2}\,.
\end{eqnarray}
As long as $\epsilon_{\rm th}$ is much smaller than $1/(m_A r_c)^2$,
it is possible to make the effective coupling $Q_{\rm eff}=
Q\,\epsilon_{\rm th} (m_A r_c)^2$ small.
However, as we will see in the Appendix, the mass $m_A$ is generally 
heavy to satisfy the condition $\epsilon_{\rm th} (m_A r_c)^2 \gg 1$
in concrete models that satisfy local gravity constraints. 
Hence this case is out of our interest.

\section{Concrete models}
\label{concrete}

In this section we consider experimental bounds on model parameters 
in concrete scalar-field potentials and investigate how these models
are consistent with local gravity constraints. 
In the previous section we have shown that the amplitude of the 
effective coupling $Q_{\rm eff}$ can be made much smaller 
than $|Q|$ in two cases:
(i) thin-shell solutions with $m_A r_c \gg 1$ and 
(ii) no-shell solutions with $m_A r_c \gg 1$.
In both cases we have found 
\begin{eqnarray}
\label{Qeff}
Q_{\rm eff} \simeq 3Q \epsilon_{\rm th}
=\frac{\phi_B-\phi_A}{2\Phi_c}\,,
\end{eqnarray}
which shows that $|Q_{\rm eff}| \ll 1$ under 
the condition $|\phi_B-\phi_A| \ll \Phi_c$.

The presence of the fifth-force interaction mediated 
by the field $\phi$ gives rise to a modification to the spherically 
symmetric metric. On using the thin-shell solution (\ref{phir}), 
the post Newtonian parameter $\gamma$ is given by 
$\gamma \simeq (1+\sqrt{6} Q_{\rm eff}/3)/
(1-\sqrt{6} Q_{\rm eff}/3)$ \cite{TUMTY}. 
The present solar-system constraint on 
$\gamma$ is $|\gamma-1|<2.3 \times 10^{-5}$ \cite{lgccon},
which comes from a time-delay effect of the Cassini tracking.
This translates into the condition 
\begin{eqnarray}
\label{bou1}
\epsilon_{\rm th, \odot}< \frac{4.7 \times 10^{-6}}{|Q|}\,,
\end{eqnarray}
where $\epsilon_{\rm th, \odot}$ is the thin-shell 
parameter for the Sun.
Under the condition $|\phi_B| \gg |\phi_A|$, the bound (\ref{bou1})
corresponds to
\begin{eqnarray}
\label{phicon1}
|\phi_{B,\odot}|<5.9 \times 10^{-11}\,,
\end{eqnarray}
where we used the value $\Phi_{c,\odot} \simeq 2.1 \times 10^{-6}$.

The fifth force induced by the field $\phi(r)$ leads to the 
acceleration of a point particle given by 
$a_{\phi}=|Q_{\rm eff} \phi(r)|$ \cite{KW2}.
This then gives rise to a difference for free-fall accelerations of 
the moon ($a_{\rm Moon}$) and the Earth ($a_{\oplus}$) toward
the Sun. Using the present experimental bound, 
$2|a_{\rm Moon}-a_{\oplus}|/(a_{\rm Moon}+a_{\oplus})<10^{-13}$ \cite{lgccon},
under the conditions that the Sun, Earth, and Moon are subject to the effect 
of reducing the effective coupling discussed above, we obtain 
the following constraint \cite{TUMTY}
\begin{eqnarray}
\label{bou2}
\epsilon_{\rm th, \oplus}< \frac{8.8 \times 10^{-7}}{|Q|}\,,
\end{eqnarray}
where $\epsilon_{\rm th, \oplus}$ is the thin-shell 
parameter for the Earth \cite{KW2}.
The thin-shell condition for the atmosphere of the Earth provides
the same order of the upper bound, provided that 
$\epsilon_{\rm th} \simeq \Delta r_c/r_c$ \cite{Capo08}.
Using the value $\Phi_{c,\oplus} \simeq 7.0 \times 10^{-10}$ for the 
Earth, the bound (\ref{bou2}) can be expressed as
\begin{eqnarray}
\label{phicon2}
|\phi_{B,\oplus}|<3.7 \times 10^{-15}\,.
\end{eqnarray}

We recall that $\phi_{B,\oplus}$ ($\phi_{B, \odot}$) depends upon 
the density far outside of the Earth (the Sun). 
For both the Earth and the Sun we take the value 
$\rho_B \simeq 10^{-24}$\,g/cm$^3$ 
that corresponds to the dark matter/baryon density in 
our galaxy. Then $|\phi_{B,\oplus}|$ is equivalent to 
$|\phi_{B,\odot}|$ provided that the scalar-field potential is specified.
In the following we adopt the severe bound (\ref{phicon2}) 
for two potentials given in Eqs.~(\ref{po1}) and (\ref{po}).

\subsection{Inverse-power law potential}

Let us consider the inverse power-law potential (\ref{po1}).
In this case we have 
\begin{eqnarray}
\label{phiB}
\phi_{B,\oplus}=\left[ \frac{n}{Q} \frac{M_{\rm pl}^4}{\rho_B}
\left( \frac{M}{M_{\rm pl}} \right)^{n+4} \right]^{\frac{1}{n+1}}M_{\rm pl}\,.
\end{eqnarray}
On using the bound (\ref{phicon2}) with $n$ and $Q$ 
of the order of unity, we get the following constraint
\begin{eqnarray}
M \lesssim 10^{-\frac{15n+130}{n+4}}M_{\rm pl}\,.
\end{eqnarray}
This shows that $M \lesssim10^{-2}$\,eV for $n=1$ and 
$M \lesssim 10^{-4}$\,eV for $n=2$, which are consistent with 
the bound derived in Ref.~\cite{KW2}.

The mass squared $m_A^2$ about the potential minimum at 
$\phi=\phi_A$ is given by 
\begin{eqnarray}
m_A^2=n(n+1) \left( \frac{n}{Q} \right)^{-\frac{n+2}{n+1}}
\left( \frac{\rho_A}{M_{\rm pl}^4} \right)^{\frac{n+2}{n+1}}
\left( \frac{M}{M_{\rm pl}} \right)^{-\frac{n+4}{n+1}}M_{\rm pl}^2\,.
\end{eqnarray}
Multiplying the squared of the radius $r_c$ of the Earth and 
eliminating $M$ with the use of Eq.~(\ref{phiB}), we obtain
\begin{eqnarray}
\label{mA}
(m_A r_c)^2=6(n+1)Q \Phi_{c,\oplus}
\left( \frac{\rho_A}{\rho_B} \right)^{\frac{1}{n+1}}
\frac{1}{\phi_{B,\oplus}}\,.
\end{eqnarray}
Since the mean density of the Earth is $\rho_A \approx 10$\,g/cm$^3$,
the experimental bound (\ref{phicon2}) leads to 
\begin{eqnarray}
\label{mcon}
m_A r_c >3 \sqrt{(n+1)Q} \cdot 10^{\frac{5n+30}{2(n+1)}}\,.
\end{eqnarray}
When $Q$ is of the order of unity one has
$m_A r_c \gtrsim 10^9$ for $n=1$ and 
$m_A r_c \gtrsim 10^7$ for $n=2$.
Hence in these cases the mass $m_A$ in fact satisfies
the condition $m_A r_c \gg 1$ so that the effective 
coupling is given by Eq.~(\ref{Qeff}).

If the body has a thin-shell then it is required that 
the condition, $\epsilon_{\rm th}>1/(m_A r_c)$, is satisfied.
Employing the relation (\ref{mA}), this 
condition corresponds to 
\begin{eqnarray}
\label{con}
m_A r_c <(n+1) \left( \frac{\rho_A}{\rho_B} \right)^{\frac{1}{n+1}}
 \simeq (n+1) \cdot 10^{\frac{25}{n+1}}\,.
\end{eqnarray}
For the compatibility of this inequality with Eq.~(\ref{mcon}), we require that 
\begin{eqnarray}
n \lesssim 4\,,
\end{eqnarray}
in which case the body has a thin-shell.
Meanwhile, when $n \gtrsim 4$, the solution corresponds to 
the no-shell case with $m_A r_c \gg 1$.

We note that the condition (\ref{con}) for the existence of thin-shells
can be also expressed as
\begin{eqnarray}
\phi_{B,\oplus}>\frac{6Q \Phi_{c,\oplus}}{n+1}
\left( \frac{\rho_B}{\rho_A} \right)^{\frac{1}{n+1}}\,.
\end{eqnarray}
When $Q={\cal O}(1)$ we have $\phi_{B,\oplus} \gtrsim 10^{-21}$ 
for $n=1$ and $\phi_{B,\oplus} \gtrsim 10^{-17}$ for $n=2$.
If future experiments provide tight bounds on $\phi_{B,\oplus}$
such as $\phi_{B,\oplus} \lesssim 10^{-21}$, 
the field profile corresponds to the no-shell case for $n \ge 1$. 
Hence the body does not have a thin-shell in such a situation.
We stress again that no-shell solutions are also consistent with 
local gravity constraints, since the effective coupling $Q_{\rm eff}$
takes the same asymptotic form as that in the thin-shell case
with the massive limit.

\subsection{The potential motivated by $f(R)$ gravity}

The next example is the potential (\ref{po}), which covers
viable $f(R)$ models that satisfy local gravity
constraints \cite{Hu,Star,Tsuji08}.
The field value $\phi_{B,\oplus}$ is given by 
\begin{eqnarray}
\label{phiB2}
|\phi_{B,\oplus}|=\frac{1}{2|Q|} 
\left( \frac{2pC V_0}{\rho_B} \right)^{\frac{1}{1-p}}M_{\rm pl}\,.
\end{eqnarray}
Note that $V_0$ is of the order of the present cosmological energy 
density $\rho_0 \simeq 10^{-29}$\,g/cm$^3$,
if the potential of the field $\phi$ is responsible for a late-time 
acceleration \cite{TUMTY}.
Employing the experimental constraint (\ref{phicon2}) with $C$ 
and $|Q|$ of the order of unity, one can derive the bound $p>14/9$.
More precise analysis using the information of the late-time acceleration 
provides the constraint \cite{TUMTY}:
\begin{eqnarray}
p>1-\frac{5}{13.8-{\rm log}_{10} |Q|}\,.
\end{eqnarray}
When $|Q|=0.1$ and $|Q|=1$ this gives
the bounds $p>0.66$ and $p>0.64$, respectively.

The quantity $(m_A r_c)^2$ for the Earth is 
\begin{eqnarray}
\label{mA2}
(m_A r_c)^2=6|Q| (1-p) \frac{\Phi_{c,\oplus}}{|\phi_{B,\oplus}|}
\left( \frac{\rho_A}{\rho_B} \right)^{\frac{1}{1-p}}\,.
\end{eqnarray}
Under the constraint (\ref{phicon2}) we get 
\begin{eqnarray}
\label{marc1}
m_A r_c \gtrsim \left[ |Q| (1-p) \right]^{1/2} \cdot 
10^{\frac{31-6p}{2(1-p)}}\,.
\end{eqnarray}
When $p>0.65$ this condition correspond to 
$m_A r_c \gtrsim 10^{39}$, which means that 
the field is extremely massive inside of the body.

The necessary condition under which the body has a thin-shell, 
$\epsilon_{\rm th}>1/(m_A r_c)$, is given by 
\begin{eqnarray}
\label{marc2}
m_A r_c < (1-p) \cdot 10^{\frac{25}{1-p}}\,.
\end{eqnarray}
It is clear that the values of $m_A r_c$ that satisfy both
(\ref{marc1}) and (\ref{marc2}) exist for $p$ ranging
in the region $0<p<1$.
Hence the body has a thin-shell under the 
current experimental bound (\ref{phicon2}) without 
reaching the no-shell regime.

The condition (\ref{marc2}) can be also expressed as 
\begin{eqnarray}
|\phi_{B,\oplus}|>\frac{6|Q| \Phi_{c,\oplus}}{1-p} 
\left( \frac{\rho_B}{\rho_A} \right)^{\frac{1}{1-p}}\,.
\end{eqnarray}
When $p=0.65$ and $|Q|=1$ one has $|\phi_{B,\oplus}| \gtrsim 10^{-79}$.
Note that this field value gets even smaller for $p$ larger than 0.65.
Since it is unlikely that future experiments reach the level of the constraint
$|\phi_{B,\oplus}| \lesssim 10^{-79}$, the field profile is well described by 
thin-shell solutions for realistic bounds on $|\phi_{B,\oplus}|$.
This situation is different from that for the inverse power-law potential.

\section{Conclusions}
\label{conclude}

In this paper we have derived analytic solutions of a chameleon 
scalar field $\phi$ in the background of a spherically symmetric body 
by taking into account a field mass $m_A$ about the potential 
minimum at $\phi=\phi_A$ inside of the body.
The mass $m_A$ is important to determine the field value 
$\phi(r_1)$ at which $\phi$ begins to evolve along the potential.
This is characterized by the condition, 
$m_A^2 \left[ \phi (r_1)-\phi_A \right]=Q \rho_A$, 
where $Q$ is a strength of the matter coupling and 
$\rho_A$ is a mean density of the body.
In the region $r_1<r<r_c$ ($r_c$ is a radius of the body), 
the presence of the matter coupling leads to a rapid evolution 
of the field. It is known that the effective matter coupling 
outside of the body gets much smaller than $|Q|$ 
if the body has a thin-shell \cite{KW,KW2}.

By considering the solution in the region $0<r<r_1$ with 
the mass $m_A$ taken into account, we have derived 
Eqs.~(\ref{ne1})-(\ref{ne3}) as 
analytic expressions of the field profile.
There exist three different regimes depending on the
position of $r_1$, i.e., (i) thick-shells ($r_1=0$), 
(ii) thin-shells ($(r_c-r_1)/r_c \ll 1$), and 
(iii) no-shells ($r_1=r_c$).
In the thick-shell case we have recovered the usual 
solution for $\phi$ in the exterior of the body, 
where the matter coupling is of the order of $Q$.
 
The standard thin-shell solution derived in Ref.~\cite{KW,KW2} 
is recovered in the limit $m_A r_c \to \infty$.
When $m_A r_c \gg 1$ the effective coupling 
outside of the body is given by 
$Q_{\rm eff} \simeq 3Q (\Delta r_c/r_c+1/m_A r_c)$
and is thus subject to the correction coming from $1/m_A r_c$.
However, if one uses the thin-shell parameter 
$\epsilon_{\rm th}=(\phi_B-\phi_A)/(6Q \Phi_c)$ introduced 
in Refs.~\cite{KW,KW2}, this effective coupling can be expressed
by $Q_{\rm eff} \simeq 3Q \epsilon_{\rm th}$.
Since the condition $\epsilon_{\rm th} \ll 1$ is used when we
place experimental bounds on model parameters, the original 
works \cite{KW,KW2} based on this criterion are 
not subject to change by the inverse mass correction.
When $m_A r_c \ll 1$ the effective coupling 
in the exterior of the body is of the order of $Q$,
which means that models with large couplings ($|Q|={\cal O}(1)$) 
are not consistent with local gravity constraints.
The requirement of the large mass ($m_A r_c \gg 1$)
for the consistency with local gravity experiments implies that 
the field needs to change rapidly from $\phi_A$ to $\phi_B$
during the transition from the thin-shell regime to the 
exterior region of the body.

In the no-shell case, the mass term $m_A^2 (\phi(r_c)-\phi_A)$
is smaller than the matter coupling term $Q\rho_A$ 
so that the field always stays around $\phi=\phi_A$ inside 
of the body. The interior and exterior solutions of the field
are given in Eqs.~(\ref{nosh1}) and (\ref{nosh2}). 
We have shown that, in the limit $m_A r_c \gg 1$, 
the field profile outside of the body and the expression of 
the effective coupling $Q_{\rm eff} \simeq 3Q \epsilon_{\rm th}$ 
are the same as those in the thin-shell case with the same massive limit. 
Since there is no lower bound for $\epsilon_{\rm th}$ contrary to 
the thin-shell case, there is a possibility that the no-shell case 
is compatible with local gravity constraints 
even in the situation that the thin-shell case is not.

We used experimental bounds coming from the violation 
of equivalence principle as well as solar-system tests to
constrain concrete models under the chameleon mechanism.
For the potential $V(\phi)=M^{4+n}\phi^{-n}$
the experimental bound (\ref{bou2}) leads
to the constraint $m_A r_c \gg 1$ for the Earth
(e.g., $m_A r_c \gtrsim 10^9$ for $n=1$ and $Q=1$).
Under the current experimental bound we have found that 
the body has a thin-shell for $n \lesssim 4$, whereas 
the field profile corresponds to no-shell solutions 
for $n \gtrsim 4$. As long as the condition, 
$M \lesssim 10^{-\frac{15n+130}{n+4}M_{\rm pl}}$, 
is satisfied, the inverse power-law potential is consistent 
with local gravity tests both for thin-shell and no-shell solutions.

We also considered another potential given in (\ref{po})
that includes viable $f(R)$ models as specific cases.
In this case the field is very massive inside of the body:
$m_A r_c \gtrsim 10^{39}$ for the values of $p$
consistent with local gravity constraints ($p \gtrsim 0.65$).
We have found that under the current experimental bound
the field profiles always correspond to thin-shell solutions
instead of no-shell solutions.
This different property compared to the inverse-power law potential 
comes from the fact that the field inside of the body 
is much heavier.

We have thus shown that the chameleon mechanism works
in a robust way provided that the field mass inside of the body 
satisfies the condition $m_A r_c \gg 1$ with appropriate
boundary conditions.
In both thin-shell and no-shell cases we just need to use the
effective coupling $Q_{\rm eff}=3Q \epsilon_{\rm th}=(\phi_B-\phi_A)/2\Phi_c$
to place experimental bounds on model parameters.
We note that under the condition $m_A r_c \gg 1$ the field 
value at $r=0$ is required to be very close to $\phi_A$, 
see Eqs.~(\ref{phi0}) and (\ref{phi0no}).
This property is especially severe for the potential (\ref{po})
due to the condition $m_A r_c \gtrsim 10^{39}$. 

It will be interesting to extend our analysis to the case in the
strong gravitational background as in the recent work \cite{Koba}
of $f(R)$ gravity (motivated by Ref.~\cite{Frolov}).
Under a strong gravity we require a careful analysis by taking into 
account the effect of gravitational potentials
to the field equation in addition to the issue of 
the boundary condition at $r=0$, 
which we leave for future work.

\section*{ACKNOWLEDGEMENTS}
S.~T. thanks financial support for JSPS (No.~30318802).

\section*{Appendix}

In this Appendix we show that no-shell solutions with a nearly massless field
does not satisfy experimental bounds for the models discussed in 
Sec.~\ref{concrete}.
In this case the effective coupling is given by 
$Q_{\rm eff}=Q \epsilon_{\rm th} (m_A r_c)^2$ from Eq.~(\ref{massless}).
The experimental test for the violation of equivalence
principle using the free-fall acceleration of the Moon and the Earth toward
the Sun provides the bound $|Q_{\rm eff}|<2.6 \times 10^{-6}$.
This translates into the condition 
\begin{eqnarray}
\label{noshellb}
|\phi_{B, \oplus}| (m_A r_c)^2<1.6 \times 10^{-5}\,\Phi_{c,\oplus}\,.
\end{eqnarray}

For the inverse power-law potential (\ref{po1}), the constraint (\ref{noshellb})
leads to
\begin{eqnarray}
\label{noshell1}
Q (n+1) \left( \frac{\rho_A}{\rho_B} \right)^{\frac{1}{n+1}}
<2.7 \times 10^{-6}\,,
\end{eqnarray}
where we used Eq.~(\ref{mA}). Since $\rho_A/\rho_B \simeq 10^{25}$
for the Earth, it is not possible to satisfy the inequality (\ref{noshell1})
for $n>0$ and $Q={\cal O}(1)$.

For the potential (\ref{po}), the constraint (\ref{noshellb}) leads to 
\begin{eqnarray}
\label{noshell2}
|Q| (1-p) \left( \frac{\rho_A}{\rho_B} \right)^{\frac{1}{1-p}}
<2.7 \times 10^{-6}\,,
\end{eqnarray}
where we used Eq.~(\ref{mA2}). 
Again we can not satisfy this condition for $0<p<1$
and $|Q|={\cal O}(1)$.


\end{document}